\documentclass{elsart}

\usepackage{graphicx}
\usepackage{epsfig}

\usepackage{amssymb}

\begin{document}

\begin{frontmatter}



\title{Charge inversion in colloidal systems}


\author{Ren\'e Messina,}
\author{Christian Holm and Kurt Kremer}

%
\address{Max-Planck-Institut f\"{u}r Polymerforschung, Ackermannweg 10, 55128 Mainz,
Germany}

\begin{abstract}
We investigate spherical macroions in the strong Coulomb coupling regime
within the primitive model in salt-free environment. Molecular dynamics (MD)
simulations are used to elucidate the effect of $discrete$ macroion charge
distribution on charge inversion. A systematic comparison is made
with the charge inversion obtained in the conventional  continuous
charge distribution.
Furthermore the effect of multivalent counterions is reported.

\end{abstract}

\end{frontmatter}

\section{Introduction}
\label{sec.intro}
Charged colloids (macroions) might exhibit very surprising
behaviors due to the Coulomb interaction and the  two different
length scales involved there: (i) microscopic (counterions) and
(ii) mesoscopic (macroions). A very spectacular effect which has
attracted great attention these last years is the phenomenon of
overcharging 
\cite{Shklowskii_PRE_1999b,Marcelo_PRE_RapCom1999,Nguyen_PRL_2000,Nguyen_JCP_2000,Messina_PRL_2000,Messina_EPL_2000,Messina_EPJE_2001,Messina_PRE_2001}. 
This situation occurs when the number of
counterions in the vicinity of the macroion surface is so high that
the macro-particle bare charge is overcompensated (charge inversion).
This phenomenon has been observed experimentally by electrophoresis
\cite{Galisteo_PCPS_1990,Walker_CS_1996}. 
On the theoretical side, a massive effort has been
recently devoted to elucidate this striking phenomenon.  Wigner
crystal theories have been successfully applied to understand
charge inversion 
\cite{Shklowskii_PRE_1999b,Nguyen_PRL_2000,Nguyen_JCP_2000,Messina_PRL_2000,Messina_EPL_2000,Messina_PRE_2001} 
and quantitative agreement was found with
simulations 
\cite{Messina_PRE_2001}.

It turns out that all numerical and analytical methods, undertaken
so far, neglect the \textit{discrete} nature of the macroion
charge distribution. But in realistic systems the macroion charge is not
continuously distributed over the macro-particle, but rather
carried by small microscopic ions. It is only very recently that a
simulation study has been reported where the macroion charge
discreteness has been considered \cite{Messina_EPJE_2001}.

In this paper, we concentrate on the effect of the macroion charge
distribution discreteness. In particular, we study this effect on
the overcharging properties. Besides we show that it is possible
to have stronger overcharge in the discrete case than in the
continuous case.

\section{Computational details}
\label{sec.comp}
\subsection{Macroion charge discretization}
\label{sec.simu-DCC}

The procedure is similar to the one used in a previous study
\cite{Messina_EPJE_2001}. The macroion charge discretization is
produced by using  $Z_{m}$ \textit{monovalent} microions of
diameter $\sigma$  distributed \textit{randomly} on the surface of
the macroion (see Fig. \ref{fig.snap}). Then the macroion structural charge
is $Q=-Z_{m}e$ \cite{CC} where $Z_{m}>0$ and $e$ is the positive
elementary charge.
The counterions have a charge
$q=+Z_{c}e$ where $Z_{c}>0$ stands for the counterion
valence. 
The discrete colloidal charges (DCC) are \textit{fixed}
on the surface of the spherical macroion and are at a distance
$r_0$ from the macroion center. Thus to produce a random discrete
charge distribution on the surface we generated (uniformly)
randomly the variables $\cos\theta$   and $\varphi$ where $\theta$ 
and $\varphi$ are the standard spherical angle coordinates.
Excluded volume is taken into account by rejecting configurations
leading to an overlap of microions.  Phenomena such as surface
chemical reactions \cite{Spalla_JCP_1991}, hydration, roughness
\cite{Bhattacharjee_Lang_1998} are not considered.

\subsection{Molecular dynamics procedure}
\label{sec.simu-MD}

A MD simulation technique was used to compute
the motion of the counterions coupled to a heat bath acting
through a weak stochastic force \textbf{W}(t).  The motion of
counterion \textit{i} obeys
\begin{equation}
\label{eq.Langevin} m\frac{d^{2}{\mathbf{r}}_{i}}{dt^{2}}= -\nabla
_{i}U({\mathbf{r}}_{i})-m\gamma
\frac{d{\mathbf{r}}_{i}}{dt}+{\mathbf{W}}_{i}(t)\: ,
\end{equation}
where \textit{m} is the counterion mass, $U$ is the
potential force having two contributions: (i) the Coulomb
interaction and (ii) the excluded volume interaction, and $\gamma$
is the friction coefficient. Friction and stochastic
force are linked by the dissipation-fluctuation theorem 
$<{{\mathbf{W}}_{i}}(t)\cdot {{\mathbf{W}}_{j}}(t')>=6m\gamma
k_{B}T\delta _{ij}\delta (t-t^{'})$. 
For the ground state
simulations the stochastic force is set to zero.

Excluded volume interactions are taken into account with a pure
repulsive Lennard-Jones potential given by

\begin{equation}
\label{eq. LJ} U_{LJ}(r)=  4 \epsilon \left[ \left( \frac{\sigma
}{r-r_{0}}\right) ^{12}- \left(\frac{\sigma}{r-r_{0}}\right)
^{6}\right] + \epsilon,
\end{equation}
%
if $r-r_0<r_{cut}$, and $U_{LJ}(r)=0$ otherwise.  For the
microion-microion interaction (the microion being either a
counterion and/or a DCC), whereas $r_{0}=7\sigma$ for the
macroion-counterion interaction, and $r_{cut} = 2^{1/6}\sigma$ is
the cutoff radius. This leads to a macroion-counterion distance of
closest approach $a=8\sigma$.

Energy and length units in our simulations are related to
experimental units by taking \( \epsilon = \)\( k_{B}T_{0} \)
(with \( T_{0}=298 \) K), and \( \sigma =3.57 \) \AA\
respectively.

The pair electrostatic interaction between any pair \textit{ij},
where \textit{i} and \textit{j} denote either a DCC, a counterion
or the central charge (for the case of the continuous surface charge distribution), reads

\begin{equation}
\label{eq.coulomb}
U_{coul}(r)=k_{B}T_{0}l_{B}\frac{Z_{i}Z_{j}}{r}\: ,
\end{equation}
where \( l_{B}=e^{2}/4\pi \epsilon _{0}\epsilon _{r}k_{B}T_{0} \)
is the Bjerrum length (here $l_B=10\sigma$) describing the electrostatic strength.

The macroion and the counterions are confined in a spherical
impenetrable cell. The macroion is held fixed
and is located at the center of the cell. To avoid image charge
complications, the permittivity $\epsilon_{r}$ is supposed to be
identical within the whole cell (including the macroion) as well
as outside the cell.

\section{Results}
\label{sec.results}

The simplest way to quantify overcharging in salt-free environment
is to compute the total electrostatic energy of the system in the
\textit{ground state} (i. e., $T=0$K) as a function of the number of overcharging
counterions $n$. 
Although this approach is very idealized, it
allows an insight into the mechanisms governing
overcharging.

\begin{figure}[hpbt]
\epsfxsize=7.5cm
\centerline{\epsfbox{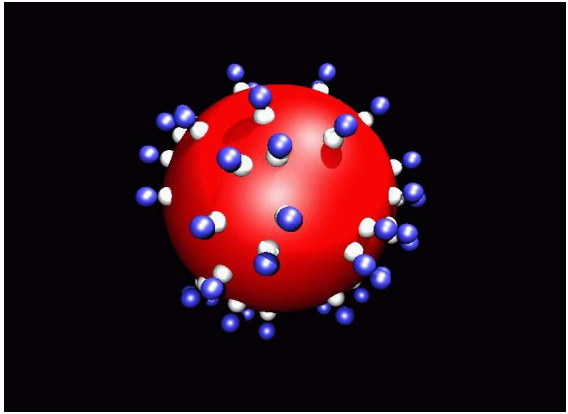}}
 \small \par{}
 \caption{
 Ground state structure for a discrete macroion charge distribution with
 $Z_m=60$.
 The monovalent discrete colloidal charges are in white and the
 monovalent counterions (here $Z_c=1$) in blue. Full ionic pairing occurs.
 }
\label{fig.snap}
\end{figure}
%

In what follows we are going to compare systematically (for a
macroion charge $Z_m=60$) the overcharging occurring in
the continuous case (CC), as is conventionally done, and
in the discrete case (with DCC ions). Fig. \ref{fig.snap} shows a typical starting
equilibrium configuration with a discrete macroion charge distribution  where the system is
globally neutral (before adding excess overcharging counterions).

\begin{figure}[hpbt]
\begin{center}
\includegraphics[width = 6.2 cm]{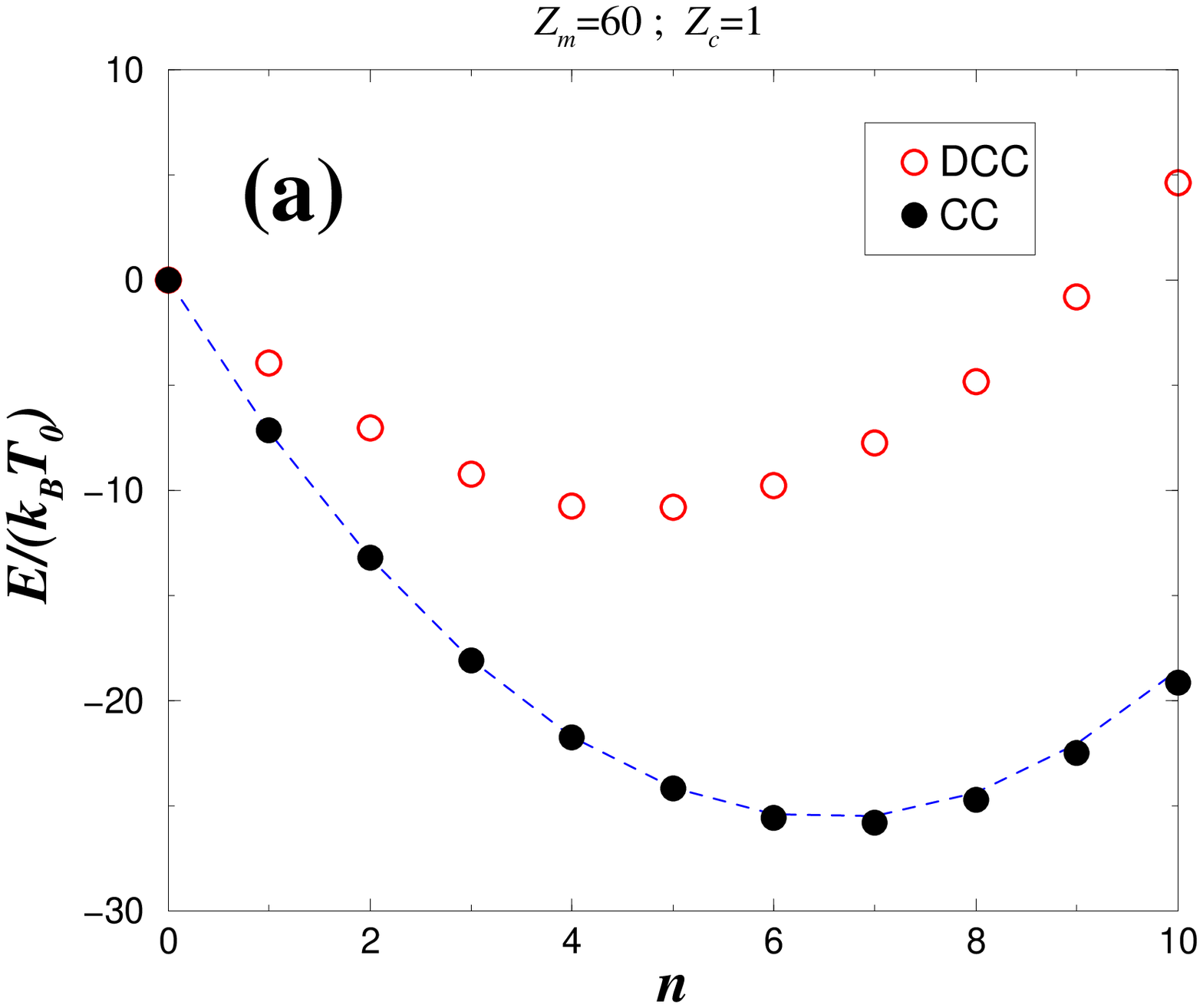}
\includegraphics[width = 6.2 cm]{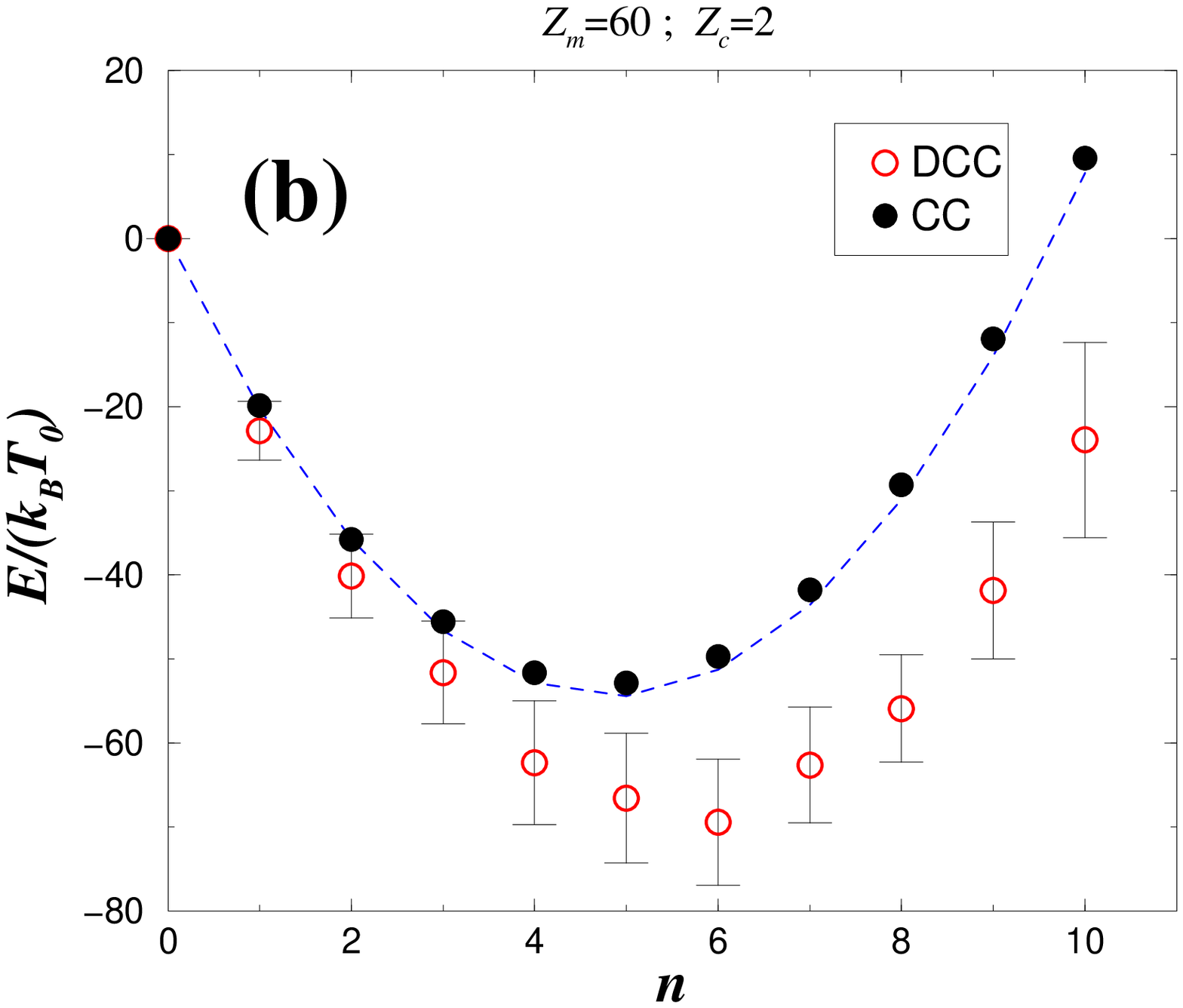}
\includegraphics[width = 6.2 cm]{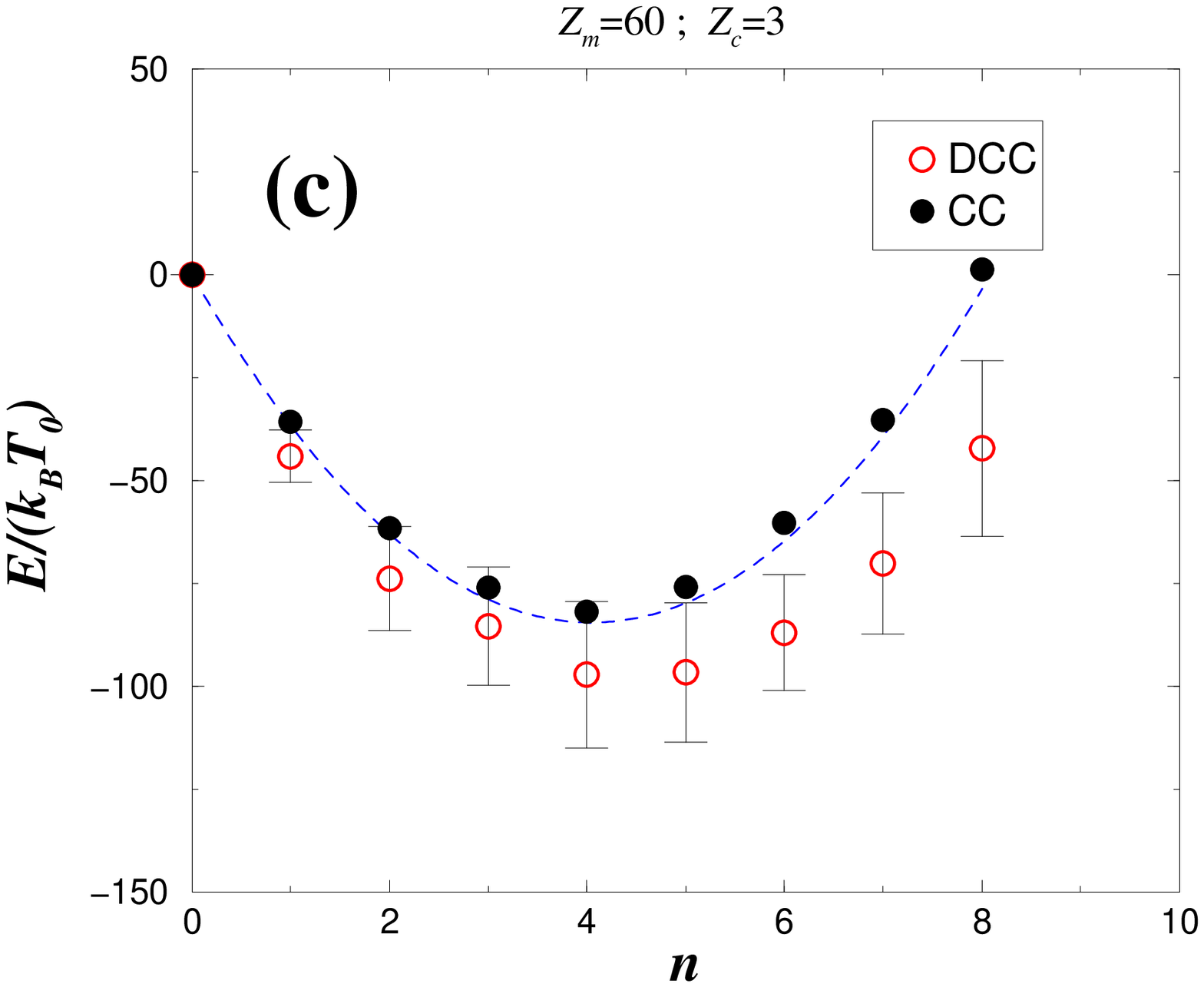}
\end{center}
\caption{
Total electrostatic energy for ground state
configurations (with $Z_m=60$) as a function of the number of
\textit{overcharging} counterions  $n$: (a) $Z_c=1$ (b)
$Z_c=2$  (c) $Z_c=3$. Overcharge curves were
computed for discrete macroion charge distribution (DCC) and
central macroionic charge (CC). The neutral case was chosen as the
potential energy origin. Dashed lines were produced by using
Wigner Crystal theory \cite{Messina_PRE_2001}. For discrete systems (DCC) error bars are
only indicated when larger than symbols.}
\label{fig.oc}
\end{figure}

Fig. \ref{fig.oc}(a) shows overcharging energy profiles for $Z_m=60$ and
$monovalent$ counterions ($Z_c=1$). It is observed that
overcharging is stronger in the continuous case than in the
discrete case as one could expect.
More precisely the gain in energy as well as the maximal number of
stabilizing overcharging counterions (corresponding to a minimum
in the $E-n$ curve) are higher when the surface macroion charge is
continuous. Indeed, having in mind that overcharging is enhanced
by counterion ordering, it is clear that in the discrete case
where strong ionic pairing occurs (between DCC site and counterion
- see Fig. \ref{fig.snap}) the counterion ordering is lowered and thus leads
to a weaker charge inversion. However since overcharging occurs in
the discrete case, the excess counterions still experience an
attractive interaction
with their neighboring dipoles (ionic pairs). This ion-dipole
interaction is the driving force for the overcharging in the discrete case 
with monovalent counterions.
Nevertheless it has been shown elsewhere \cite{Messina_EPJE_2001} that by decreasing the mean separation
between DCC sites (i. e., increasing macroion charge density) the difference
in overcharging between discrete and continuous macroion charge distribution decreases.

When we consider $multivalent$ counterions the situation is qualitatively
different. Indeed, one observes that for $Z_c=$ 2 and 3 overcharging is
even stronger in the discrete case [see Figs. \ref{fig.oc} (b-c)].
This is a rather counter-intuitive 
phenomenon, since ionic pairing is still effective. 
The crucial difference
between monovalent and multivalent counterion systems is that overcharging 
occurs with ionic pairing for multivalent counterions (since free monovalent DCC sites
are unbound), whereas for monovalent counterions only ion-dipole interaction
can take place (since in the neutral state $each$ DCC site is already bound with a
monovalent counterion). 
Obviously for these systems under
consideration depicted in Figs. \ref{fig.oc} (b-c), the ionic pairing energy is a
fundamental quantity that controls the overcharging strength. 
More precisely, the stronger the ionic bonding the more efficiently it  overcomes the
$excess$ overcharge "self-energy", and therefore the higher the overcharging
for multivalent counterion systems. 
  
Beside of the ionic bonding strength there is another
important parameter controlling overcharging: the surface macroion charge density. 
When the macroion charge density
is high enough (not reported here) then the $E(n)$ curves become quasi-identical
for discrete and continuous charge distributions. In terms of length scales,
the relevant parameter controlling charge reversal with discrete distribution is
the ration $\rho$ between the mean distance between DCC and the distance
separation of the ionic pair. For low $\rho$ the effects of discretization 
are very important whereas for $\rho\rightarrow 1$ one recovers the continuous limit.

\section{Conclusions}
We carried out MD simulations to elucidate the effect of discrete macroion
charge distribution on the overcharging. We showed that overcharging is still
effective in the discrete case. For multivalent counterions and a
sufficiently low macroion charge density, overcharging can even be stronger in
the discrete case than in the continuous case. Future works will address aqueous
solutions at finite temperature as well as the presence of salt-ions. 


\begin{thebibliography}{10}
\bibitem{CC}
Note that in the usual $continuous$ case (and
only in this case), thanks to Gauss' theorem it is equivalent (as far
as the electrostatic field outside the macroion is concerned) to consider 
a central point charge carrying the charge $Q$.
\end{thebibliography}







\end{document}